\documentstyle[12pt,twoside]{article}
\normalbaselines
\makeatletter
\@addtoreset{equation}{section}
\makeatother

\newcommand{\be}{\begin{equation}}
\newcommand{\ee}{\end{equation}}

\title{Distorted Heisenberg Algebra and Coherent States for
Isospectral Oscillator Hamiltonians}

\bigskip \bigskip

\author{David J. Fern\'andez C.${}\sp1$\thanks{e-mail:
        david@fis.cinvestav.mx},
        Luis M. Nieto${}\sp2$\thanks{e-mail:
                                     lmnieto@cpd.uva.es}
        and Oscar Rosas-Ortiz${}\sp1$\thanks{e-mail:
                                     orosas@fis.cinvestav.mx}
\\ \\
      ${}\sp1$\small {\it Departamento de F\'\i sica,
                                           CINVESTAV-IPN,}\\
      \small {\it A.P. 14-740, 07000 M\'exico D.F., Mexico}\\
              \\
      ${}\sp2$\small{\it Departamento de F\'{\i}sica Te\'orica,
                       Universidad de Valladolid}\\
      \small {\it 47011 Valladolid, Spain}}

\bigskip \bigskip

\begin{document}

\maketitle

\thispagestyle{empty}

\begin{abstract}
The dynamical algebra associated to a family of isospectral oscillator
Hamiltonians is studied through the analysis of its representation in
the basis of energy eigenstates. It is shown that this representation
becomes similar to that of the standard Heisenberg algebra, and it is
dependent of a parameter $w\geq 0$. We name it {\it distorted
Heisenberg algebra}, where $w$ is the distortion parameter. The
corresponding coherent states for an arbitrary $w$ are derived, and
some particular examples are discussed in full detail. A prescription
to produce the squeezing, by adequately selecting the initial state of
the system, is given.
\end{abstract}

\bigskip\bigskip

{\footnotesize
\begin{description}
\item[Key-Words:] Coherent states, Heisenberg algebra, squeezed
states
\item[PACS:] 03.65.-w, 11.30.Pb, 42.50.Dv

\end{description}}

\vfill

\hfill {\bf CIEA-GR-9501}

\newpage

\baselineskip=18pt
\setcounter{page}{1}
\section{Introduction}\label{sec1}

The well known coherent states of the harmonic oscillator turned out
one of most useful tools of quantum theory \cite{ks,pe,zf}. Introduced
long ago by Schr\"odinger \cite{sc}, they were employed later on by
Glauber and other authors in quantum optics \cite{gl,su,kl}. Further
developments of the subject made possible to set up some specific
definitions, applicable to various physical systems.

One possibility is to define the coherent states as eigenstates of an
annihilation operator. Following this idea, the coherent states for a
family of Hamiltonians isospectral to the harmonic oscillator were
recently derived \cite{fh}. As there is a certain arbitrariness in the
selection of the annihilation and creation operators for that system,
the most obvious realization was chosen: the operators are adjoint to
each other but their commutator is not the identity. In the same paper
a different option of constructing the lowering and rising operators
was also pursued: the creator was altered while the annihilator was
not; they were not adjoint to each other anymore, but their commutator
was equal to the identity. This modified pair, in principle, could
induce new coherent states, consistent with the application of a
``displacement operator" to the extremal state. However, the states so
derived turned out to be identical to the ones previously defined as
eigenstates of the annihilator.

In the light of those results, it is interesting to pose the following
questions: can both ideas be unified to yield lowering and rising
operators which would be adjoint to each other and would commute to
the identity, imitating then the Heisenberg algebra? If so, what kind
of coherent states would they generate?

The goal of this paper is to find out the answers. In Section 2 we
will sketch the derivation of the family of Hamiltonians isospectral
to the harmonic oscillator \cite{mi,fh}. Section 3 contains the
construction of new couples of annihilation and creation operators for
those Hamiltonians; we will build those couples from the generators of
the standard Heisenberg algebra. Indeed, we will see that there is a
family of such a pairs depending on a parameter $w\geq 0$. In Section
4 two sets of coherent states will be found for arbitrary values of
$w$: the ones derivable as eigenstates of the annihilation operator
and the ones resulting from the application of a ``displacement"
operator on the extremal state. By fixing some specific values of $w$,
we will attain three particularly interesting cases which will be
discussed in Section 5. We conclude with some general remarks in
Section 6.

\section{The isospectral oscillator Hamiltonians
$H\sb\lambda$}\label{sec2}

We are interested in a family of Hamiltonians $H\sb\lambda$ which can
be derived from the harmonic oscillator Hamiltonian $H$ using a
variant of the factorization method \cite{mi}. The standard
factorization expresses $H$ as two products
\be
H=aa\sp\dagger -\frac 12, \qquad H=a\sp\dagger a+\frac{1}{2},
\ee
where $H$ and the annihilation $a$ and creation $a\sp\dagger$
operators are given by
\be
H=-\frac 12\frac{d\sp2}{dx\sp2}+\frac{x\sp2}{2},  \quad
a=\frac{1}{\sqrt{2}}\left(\frac{d}{dx}+x\right), \quad
a\sp\dagger=\frac{1}{\sqrt{2}}\left(-\frac{d}{dx}+x\right).
\ee
It can be proved that the first decomposition in (2.1) is not unique.
Indeed, there exist more general operators $b$ and $b\sp\dagger$
generating $H$:
\be
H=bb\sp\dagger-\frac 12 ,  \quad
b=\frac{1}{\sqrt{2}}\left(\frac{d}{dx}+\beta(x)\right), \quad
b\sp\dagger=\frac{1}{\sqrt{2}}\left(-\frac{d}{dx}+\beta(x)\right).
\ee
Hence, $\beta(x)$ obeys the Riccati equation
$\beta'+\beta\sp2=x\sp2+1$, whose general solution is
\be
\beta(x)=x+\frac{e\sp{-x\sp2}}{\lambda+\int\sb0\sp x e\sp{-
y\sp2}dy},\quad \lambda\in{\bf R}.
\ee
The inverted product $b\sp\dagger b$ is not related to $H$, but
induces a different Hamiltonian
\begin{eqnarray}
H\sb\lambda & = & b\sp\dagger b+\frac 12=-\frac 12 \,
\frac{d\sp2}{dx\sp2}+V\sb\lambda(x), \\    V\sb\lambda(x) & = &
\frac{x\sp2}{2}-\frac{d}{dx}\left[\frac{e\sp{-
x\sp2}}{\lambda+\int\sb0\sp x e\sp{-y\sp2} dy}\right],\quad
\vert\lambda\vert>\frac{\sqrt{\pi}}{2}.
\end{eqnarray}
The Hamiltonians $H$ and $H\sb\lambda$ are connected by the following
relation:
\be
H\sb\lambda b\sp\dagger = b\sp\dagger (H+1).
\ee
Therefore, if $\vert\psi\sb{n}\rangle$ are the standard eigenstates of
$H$ verifying $H\vert\psi\sb{n}\rangle
=(n+1/2)\vert\psi\sb{n}\rangle$, the states defined as
\be
\vert\theta\sb n\rangle = \frac{b\sp\dagger\vert\psi\sb{n-
1}\rangle}{\sqrt{n}},\ \ n = 1,2,3,\cdots
\ee
are normalized orthogonal eigenstates of $H\sb\lambda$ with
eigenvalues $E\sb n=n+1/2$ respectively. The ground state of
$H\sb\lambda$ is disconnected from the other eigenstates, it has
eigenvalue $E\sb0=1/2$ and satisfies $b\vert\theta\sb0\rangle =0$. In
the coordinate representation it is given by
\be
\theta\sb0(x) \propto \frac{e\sp{-x\sp2/2}}{\lambda+\int\sb0\sp x
e\sp{-y\sp2} dy}.
\ee
Summarizing this section, $\lbrace  H\sb\lambda,\
\vert\lambda\vert>\sqrt{\pi}/2\rbrace $ represents a family of
Hamiltonians with the same spectrum as the harmonic oscillator. The
relations necessary to set up the creation and annihilation operators
of $H\sb\lambda$ are
\begin{eqnarray}
b \vert\theta\sb n\rangle & = & \sqrt{n}\,\vert\psi\sb{n-1}\rangle,
\qquad b\sp\dagger\vert\psi\sb
n\rangle=\sqrt{n+1}\,\vert\theta\sb{n+1}\rangle, \nonumber \\ a
\vert\psi\sb n\rangle & = & \sqrt{n}\,\vert\psi\sb{n-1}\rangle,
\qquad a\sp\dagger\vert\psi\sb
n\rangle=\sqrt{n+1}\,\vert\psi\sb{n+1}\rangle.
\end{eqnarray}

\section{Distorted Heisenberg algebra of $H\sb\lambda$}\label{sec3}

It is important to identify now a suitable pair of annihilation and
creation operators for $H\sb\lambda$. The obvious choice follows
immediately from (2.10) \cite{mi,fh}:
\be
A=b\sp\dagger ab, \qquad A\sp\dagger=b\sp\dagger a\sp\dagger b.
\ee
The effective action of, let us say, the annihilation operator $A$
comes after three intermediate transformations: we take an eigenstate
$\vert\theta\sb n\rangle$ of $H\sb\lambda$ and transform it, by the
action of $b$, in $\vert\psi\sb{n-1}\rangle$, an eigenstate of $H$;
then, $a$ transforms $\vert\psi\sb{n- 1}\rangle$ in $\vert\psi\sb{n-
2}\rangle$; finally, $\vert\theta\sb{n-1}\rangle$ is obtained through
the action of $b\sp\dagger$ on $\vert\psi\sb{n-2}\rangle$. A similar
procedure works for $A\sp\dagger$.

As it is easily seen, the operators defined in (3.1) are reciprocally
adjoint, but they do not commute to the identity. Different
annihilation and creation operators arise if $A$ is left unchanged but
we define a new creator $B\sp\dagger$, with the requirement
$[A,B\sp\dagger]=1$. The operator $B\sp\dagger$ turns out to be
\cite{fh}
\be
B\sp\dagger =b\sp\dagger\, a\sp\dagger\, \frac{1}{(N+1)(N+2)}\, b.
\ee
Obviously, $B\sp\dagger$ is not the adjoint of $A$. A third
realization, and this is one of the results of this paper, arises when
both $A$ and $A\sp\dagger$ are substituted by new annihilation and
creation operators $C$ and $C\sp\dagger$ chosen to be reciprocally
adjoint, and such that their commutator is the identity on a subspace
${\cal H}\sb s$ of the state space ${\cal H}$:
\be
[C,C\sp\dagger]=1 \quad {\rm on} \quad {\cal H}\sb s \subset {\cal H}.
\ee
In the spirit of \cite{fh}, we propose
\be
C=b\sp\dagger f(N) ab, \qquad C\sp\dagger=b\sp\dagger a\sp\dagger
f(N)b,
\ee
where $f(x)$ is a real function to be determined, and $N=a\sp\dagger
a$ is the standard number operator. Taking into account (2.10),
$[C,C\sp\dagger]$ acts on $\vert\theta\sb n\rangle$ as follows
\be
[C,C\sp\dagger]\vert\theta\sb n\rangle = \left\lbrace  n\sp2(n+1)[f(n-
1)]\sp2-(n-1)\sp2 n [f(n- 2)]\sp2 \right\rbrace \vert\theta\sb
n\rangle.
\ee
For $n=0$ and $n=1$ we get
\be
[C,C\sp\dagger]\vert\theta\sb0\rangle=0, \qquad
[C,C\sp\dagger]\vert\theta\sb1\rangle = 2[f(0)]\sp2
\vert\theta\sb1\rangle.
\ee
We impose now (3.3), with ${\cal H}\sb s$ the subspace generated by
$\lbrace \vert \theta\sb n \rangle; n\geq 2\rbrace $. Defining a new
function, $g(x)=(x+1)\sp2(x+2)[f(x)]\sp2, \ x\in {\bf N}$, it has to
verify the difference equation
\be
g(x+1)-g(x)=1,
\ee
whose general solution is well known \cite{da}
\be
g(x)=x+w(x),
\ee
with $w(x)$ obeying
\be
w(x+1)=w(x) \quad \forall \ x\in{\bf N};
\ee
in other words, $w(x)$ is an arbitrary periodic function, with period
equal to one. Therefore, $f(x)$ takes the form
\be
f(x)=\frac{1}{x+1}\sqrt{\frac{x+w(x)}{x+2}}.
\ee
The function $f(x)$ must be real, hence, $x+w(x)\geq 0, \ \forall \
x\in{\bf N}$. This fact, and the periodicity of $w(x)$ imply that the
relevant value of $w(x)$ is $w(0)$ with
\be
w\equiv w(0)\geq 0.
\ee
The form of the operators $C$ and $C\sp\dagger$ satisfying (3.3) is
then
\be
C\sb w=b\sp\dagger\frac{1}{N+1}\sqrt{\frac{N+w(N)}{N+2}} ab, \qquad
C\sb w\sp\dagger=b\sp\dagger
a\sp\dagger\frac{1}{N+1}\sqrt{\frac{N+w(N)}{N+2}} b,
\ee
where the subindex labels the dependence of $C$ and $C\sp\dagger$ on
the parameter $w$. The action of $C\sb w, \ C\sb w\sp\dagger$ and
$[C\sb w,C\sb w\sp\dagger]$ on $\lbrace  \vert\theta\sb n\rangle, \
n\in{\bf N}\rbrace $ is
\begin{eqnarray}
C\sb w\vert\theta\sb n\rangle & = & (1-\delta\sb{n,0}-\delta\sb{n,1})\
\sqrt{n-2+w}\ \vert\theta\sb{n-1}\rangle,  \\ C\sb w\sp\dagger
\vert\theta\sb n\rangle & = & (1-\delta\sb{n,0})\ \sqrt{n-1+w}\
\vert\theta\sb{n+1}\rangle ,
\end{eqnarray}
\be
[C\sb w,C\sb w\sp\dagger]\vert\theta\sb n\rangle=  [1-\delta\sb{n,0} +
\delta\sb{n,1}(w -1)] \vert\theta\sb n\rangle =\left\lbrace
\begin{array}{ll}  0, & n=0 ;\\ w  \vert\theta\sb1\rangle , & n=1 ;\\
\vert\theta\sb n\rangle, &n \geq 2 . \end{array} \right.
\ee
As this action resembles that of the generators of the Heisenberg
algebra $a, \ a\sp\dagger$ and $[a,a\sp\dagger]$ on the harmonic
oscillator basis $\lbrace \vert\psi\sb n\rangle\rbrace $, we are led
to define the two operators
\be
X\sb w=\frac{C\sp\dagger\sb w + C\sb w}{\sqrt{2}},\qquad P\sb w=i\
\frac{C\sp\dagger\sb w - C\sb w}{\sqrt{2}}.
\ee
They are to $H\sb\lambda$ as the usual coordinate $X$ and momentum $P$
operators are to the harmonic oscillator Hamiltonian. The commutator
of $X\sb w$ and $P\sb w$ in the basis $\lbrace \vert\theta\sb
n\rangle, \ n\in{\bf N}\rbrace $ is
\be
[X\sb w,P\sb w]\vert\theta\sb n\rangle=  i[1-\delta\sb{n,0} +
\delta\sb{n,1}(w -1)] \vert\theta\sb n\rangle =\left\lbrace
\begin{array}{ll}  0, & n=0 ;\\ iw  \vert\theta\sb1\rangle , & n=1 ;\\
i\vert\theta\sb n\rangle, &n \geq 2 . \end{array} \right.
\ee
{}From equations (3.13)-(3.17), it can be seen that the representation
of $C\sb w, \ C\sb w\sp\dagger$,  $X\sb w$ and $P\sb w$ on the basis
$\lbrace \vert\theta\sb n \rangle,\ n\in{\bf N}\rbrace $ is reducible
because there are two invariant subspaces, one of them generated by
$\vert\theta\sb0\rangle$ and the other one by $\lbrace \vert\theta\sb
n \rangle, n\geq1\rbrace $. We denote them as ${\cal H}\sb0$ and
${\cal H}\sb r$, respectively. In ${\cal H}\sb0$ all the operators
$C\sb w, \ C\sb w\sp\dagger$, $X\sb w$ and $P\sb w$ are trivially
represented by the $1\times 1$ null matrix. The relevant
representation for those operators arises when we consider their
action on vectors $\vert\psi\rangle\in {\cal H}\sb r$. This
representation is similar to the one of the standard Heisenberg
algebra; however, it depends on the parameter $w$. That makes the
difference between the two representations here compared. Thus, we
call ``distorted Heisenberg algebra'' the algebra generated by $C\sb
w$ and $C\sb w\sp\dagger$ (or by $X\sb w$ and $P\sb w$). One reason to
choose this name is because the representation of $C\sb w$ and $C\sb
w\sp\dagger$ on ${\cal H}\sb r$ can be though of as coming from that
of $a$ and $a\sp\dagger$ on ${\cal H}$ after a two steps distortion:
first, we remove the ground state of the oscillator Hamiltonian;
second, we deform the representation induced by the remaining basis
vectors through the introduction of a distortion parameter $w$.
However, it is important to recall that $C\sb w, \ C\sb w\sp\dagger, \
X\sb w$ and $P\sb w$ are not simple generalizations of $a, \
a\sp\dagger, \ X$ and $P$, in the sense that it is impossible to get
the action of the last ones on ${\cal H}$ as a limit procedure, for
$w$ tending to a specific value, of the action of the first ones on
${\cal H}\sb r$. We postpone to Section 5 the discussion of cases, for
particular values of $w$, for which the distorted Heisenberg algebra
is the closest to the Heisenberg algebra on ${\cal H}$. This will give
a better support to our terminology. Meanwhile, here and in the next
section we  derive the general results, valid for the full range
$w\geq 0$.

\section{New coherent states of $H\sb\lambda$}\label{sec4}

It is well known that, for a general system, there are three
non-equivalent definitions for the coherent states \cite{ks,pe,zf}.
One consists in defining them as eigenstates of the annihilation
operator of the system, denoted here by $J$. Another possibility is to
define them as the vectors resulting from the application of the
unitary displacement operator $\exp(zJ\sp\dagger - \bar zJ)$ on an
extremal state $\vert\varphi\sb0 \rangle$, which usually is an
eigenstate of $J$ with zero eigenvalue, i.e., $J\vert\varphi\sb0
\rangle=0$ (here $z\in{\bf C}$, the bar over $z$ means complex
conjugation, and $J\sp\dagger$ denotes the adjoint of $J$). A third
definition characterizes the coherent states as minimum-uncertainty
states (see also \cite{ns}). In \cite{fh} a set of coherent states for
$H\sb\lambda$ was derived  using the first definition, with the
annihilation operator $A$ given by (3.1). An additional set of
coherent states was found through the second definition, but employing
the non-unitary displacement operator $\exp(zB\sp\dagger-\bar z A)$,
the extremal state $\vert\theta\sb1\rangle$, and the operator
$B\sp\dagger$ given in (3.2). The two sets turned out to be equal
\cite{fh}.

In this section, new coherent states of $H\sb\lambda$ will be
constructed using the first definition and a modified version of the
second one departing from the annihilation and creation operators
given in (3.12). In both cases, the uncertainty product of the
distorted position and momentum operators of (3.16) on the resultant
states will be found in order to compare our new coherent states and
the standard ones, which minimize the uncertainty product $(\Delta
X)(\Delta P)$.


\subsection{Coherent states as eigenstates of $C\sb w$}

Let us denote the coherent states  $\vert z,w\rangle$, to show
explicitly their dependence on the parameter $w$. They are eigenstates
of $C\sb w$:
\be
C\sb w\vert z,w\rangle=z\vert z,w\rangle,\qquad z\in{\bf C}.
\ee
To have their explicit form, we decompose $\vert z,w\rangle$ in terms
of the basis $\vert\theta\sb n\rangle$:
\be
\vert z,w\rangle = \sum\sb{n=0}\sp{\infty} a\sb n \ \vert \theta\sb
n\rangle.
\ee
Substituting this expression in (4.1) and using (3.13) we get the
coefficients $a\sb n$. If we suppose that $w\neq0$, we get
\be
a\sb0=0, \qquad a\sb{n+1}=\frac{\sqrt{\Gamma(w)} \ z\sp
n}{\sqrt{\Gamma(w+n)}}\, a\sb1, \quad n\in{\bf N}.
\ee
If we chose $a\sb1\geq0$, the normalization condition leads to
\be
\vert z,w\rangle =\sqrt{\frac{\Gamma(w)}{{}\sb1F\sb1(1,w;r\sp2)}}\
\sum\sb{n=0}\sp{\infty} \frac{z\sp n}{\sqrt{\Gamma(w+n)}} \ \vert
\theta\sb{n+1}\rangle , \quad z=re\sp{i\varphi},
\ee
where ${}\sb1F\sb1(a,b;x)$ is the hypergeometric function
\be
{}\sb1F\sb1(a,b;x)= \frac{\Gamma(b)}{\Gamma(a)}\
\sum\sb{k=0}\sp{\infty}\frac{\Gamma(a+k)}{\Gamma(b+k)}\ \frac{x\sp
k}{k!},
\ee
$r=\vert z\vert\in{\bf R}$, and $\varphi\in{\bf R}$. We realize again,
like in \cite{fh}, that $z=0$ is doubly degenerated with eigenkets
$\vert \theta\sb0\rangle$ and $\vert \theta\sb1\rangle$.

Observe that, although the case $w=0$ is excluded of (4.3), the states
$\vert z,w\rangle$ of (4.4) tend to a well-defined limit when
$w\rightarrow 0$
\be
\vert z,0\rangle =e\sp{i\varphi} e\sp{-r\sp2/2}\
\sum\sb{n=0}\sp{\infty} \frac{z\sp n}{\sqrt{n!}} \ \vert
\theta\sb{n+2}\rangle.
\ee
We have checked this result by performing a direct calculation,
similar as the previous one, but taking $w=0$ from the very beginning,
which led us to the same states (4.6) (modulo a phase). These will be
considered again in Section 5.

Let us analize the completeness of the set $\lbrace \vert
\theta\sb0\rangle$, $\vert z,w\rangle$;  $ z\in {\bf C}\rbrace$ . We
impose
\begin{eqnarray}
I&=&\vert \theta\sb{0}\rangle\langle\theta\sb{0}\vert + \int \vert z,w
\rangle\langle z,w\vert  \, d\mu(z,w) \nonumber\\ & = & \vert
\theta\sb{0}\rangle\langle\theta\sb{0}\vert + \int \vert
z,w\rangle\langle z,w\vert \, \sigma(r,w)\, r \, dr\, d\varphi,
\end{eqnarray}
where $d\mu(z,w)$ is the unknown measure. Following a standard
procedure \cite{bg,bd}\ one finds
\be
\sigma(r,w)= \frac{{}\sb1F\sb1(1,w;r\sp2)}{\pi\Gamma(w)}\ e\sp{-
r\sp2}\ r\sp{2(w -1)} .
\ee
Here, $\sigma(r,w)$ is simpler than the corresponding function
obtained in \cite{fh}. It is possible to express any coherent states
$\vert z',w\rangle$  in terms of the others:
\be
\vert z',w\rangle=\int\vert z,w \rangle\langle z,w\vert z',w\rangle
d\mu(z,w) , \quad \vert z',w\rangle\neq \vert \theta\sb0\rangle,
\ee
with kernel given by
\be
\langle z,w \vert z',w \rangle = \frac{\sb1F\sb1(1,w; \bar z
z')}{\sqrt{\sb1F\sb1(1,w;r\sp2)  \ {}\sb1F\sb1(1,w;r'\sp2)}} .
\ee
Using (4.7), any element $\vert h\rangle\in {\cal H}$ can be expanded
in terms of the coherent states as
\be
\vert h\rangle=h\sb0 \vert \theta\sb0\rangle+\int {\tilde h} (z,\bar z
,w) \vert z,w\rangle d\mu(z,w),
\ee
where $h\sb0\equiv \langle\theta\sb0\vert h\rangle$, and
\be
{\tilde h} (z,\bar z, w)\equiv \langle z,w\vert h\rangle=
\sqrt{\frac{\Gamma(w)}{{}\sb1F\sb1(1,w;r\sp2)}}\
\sum\sb{n=0}\sp{\infty} \frac{\bar z\sp n}{\sqrt{\Gamma(w+n)}}
 \langle\theta\sb{n+1}\vert h\rangle.
\ee
The time evolution of $\vert z,w\rangle$ is quite simple:
\be
U(t)\vert z,w\rangle =
\sqrt{\frac{\Gamma(w)}{{}\sb1F\sb1(1,w;r\sp2)}}\
\sum\sb{n=0}\sp{\infty} \frac{z\sp n}{\sqrt{\Gamma(w+n)}} e\sp{-
itH\sb\lambda} \vert \theta\sb{n+1}\rangle = e\sp{-i3t/2} \vert
z(t),w\rangle,
\ee
where $z(t)\equiv z e\sp{-it}$, and $U(t)$ is the evolution operator
of the system from $0$ to $t$.

The mean value and the uncertainty of an operator $K$ in the coherent
state $\vert z,w\rangle$ are denoted
\be
\langle K\rangle\equiv \langle z,w\vert K\vert z,w\rangle, \qquad
\Delta K\equiv\sqrt{\langle {K\sp2}\rangle-\langle K\rangle\sp2}.
\ee
For the Hamiltonian $H\sb\lambda$ we get
\be
\langle H\sb\lambda \rangle = \frac12 + \frac{ {}\sb1F\sb1(2,w; r\sp2)
}{ {}\sb1F\sb1(1,w; r\sp2) } .
\ee
By the previous construction, $\langle C\sb w\rangle=z$ and $\langle
C\sb w\sp\dagger\rangle=\bar z$. Therefore, $\langle X/\sb w\rangle$
and  $\langle P\sb w\rangle$ become
\be
\langle X\sb w\rangle=\frac{ \langle C\sb w\sp\dagger\rangle+\langle
C\sb w\rangle }{\sqrt 2}=\frac{\bar z+z}{\sqrt 2} ,
\ee
\be
\langle P\sb w\rangle=i \frac{ \langle C\sb w\sp\dagger\rangle-\langle
C\sb w\rangle } {\sqrt 2}=i \frac{\bar z-z}{\sqrt 2} .
\ee
They are equal to the corresponding harmonic oscillator results. Let
us calculate now
\be
\langle C\sb w C\sb w\sp\dagger+C\sb w\sp\dagger C\sb w\rangle=w-1 +
r\sp2 + \frac{ {}\sb1F\sb1(2,w; r\sp2) } { {}\sb1F\sb1(1,w; r\sp2)
},\quad \langle C\sb w\sp2\rangle=\overline{\langle C\sb w\sp{\dagger
2}\rangle}=z\sp2.
\ee
Hence,
\begin{eqnarray}
\langle X\sb w\sp2\rangle & = & \frac12(z\sp2+\bar z\sp2)+\frac{w-1}2
+ \frac12 \left( r\sp2+ \frac{ {}\sb1F\sb1(2,w;r\sp2)}{
{}\sb1F\sb1(1,w;r\sp2)}\right), \\ \langle P\sb w\sp2\rangle & = & -
\frac12(z\sp2+\bar z\sp2)+\frac{w-1}2 + \frac12 \left(r\sp2+ \frac{
{}\sb1F\sb1(2,w;r\sp2)}{ {}\sb1F\sb1(1,w;r\sp2)}\right).
\end{eqnarray}
Thus, the uncertainties of $X\sb w$ and $P\sb w$, and their product,
are
\be
(\Delta X\sb w)\sp2 = (\Delta P\sb w)\sp2= (\Delta X\sb w)(\Delta P\sb
w) = \frac12 \left(w - 1 - r\sp2 + \frac{ {}\sb1F\sb1(2,w; r\sp2) }{
{}\sb1F\sb1(1,w;r\sp2) }\right) .
\ee
Notice that the uncertainty relation (4.21) has radial symmetry on the
complex plane of $z$. A plot of it as a function of $r=\vert z\vert$
for different values of  $w$ is shown in Figure 1. As we can see, $w/2
\leq(\Delta X\sb w)(\Delta P\sb w)\leq 1/2$ if $0\leq w\leq 1$ and
$1/2 \leq(\Delta X\sb w)(\Delta P\sb w)\leq w/2$ if $w\geq 1$. Thus,
the coherent states just derived are close to the  minimum uncertainty
ones for large values of $r$, and also for small values of $w$. In
Section 5 we will find explicit values of $w$ for which our coherent
states become minimum uncertainty states satisfying $(\Delta X\sb
w)(\Delta P\sb w) = 1/2$ for any $z$.

One question arises naturally: what happens in the harmonic oscillator
limit? This can be answered if we realize that, in the limit $\vert
\lambda \vert \rightarrow\infty$, $H\sb\lambda \rightarrow H$.
Moreover, in this limit
\be
b\rightarrow a, \quad b\sp\dagger\rightarrow a\sp\dagger, \quad \vert
\theta\sb n\rangle\rightarrow\vert \psi\sb n\rangle.
\ee
Therefore, the corresponding limits for $C\sb w$ and $C\sb
w\sp\dagger$ are
\be
C\sb{w,0}\equiv\lim\sb{\vert \lambda\vert \to\infty}C\sb w
=a\sp\dagger \frac1{N+1}\, \sqrt{\frac{N+w(N)}{N+2}}\, a\sp2,
\ee
\be
C\sb{w,0}\sp\dagger\equiv\lim\sb{\vert \lambda\vert \to\infty}C\sb
w\sp\dagger =a\sp{\dagger 2} \frac1{N+1}\, \sqrt{\frac{N+w(N)}{N+2}}\,
a .
\ee
For the coherent states we have
\be
\vert z,w\rangle\sb{0}  \equiv\lim\sb{\vert \lambda\vert
\to\infty}\vert z,w\rangle =
\sqrt{\frac{\Gamma(w)}{{}\sb1F\sb1(1,w;r\sp2)}}\
\sum\sb{n=0}\sp{\infty} \frac{z\sp n}{\sqrt{\Gamma(w+n)}} \ \vert
\psi\sb{n+1}\rangle.
\ee
We see that, in general, the coherent states derived here are
different from the standard ones of $H$, even though
$H\sb\lambda\rightarrow H$ when $\vert\lambda\vert\rightarrow\infty$.
In Section 5, we will analyse other limit cases, by approaching
specific values of $w$, which will provide us with more insight about
the differences and similarities of our coherent states and the
standard ones.

\subsection{Displacement operator technique and coherent states}

According to the second definition, the coherent states we should find
now would result from the application of the displacement operator
$D(z)=\exp(zC\sb w\sp\dagger-\bar z C\sb w)$ on an extremal state
$\vert\varphi\sb0\rangle$ which obeys $C\sb
w\vert\varphi\sb0\rangle=0$. For $H\sb\lambda$ and $C\sb w$ given by
(2.5), (2.6) and (3.12) there are two extremal states
$\vert\theta\sb0\rangle$ and $\vert\theta\sb1\rangle$:
\be
C\sb w\vert\theta\sb1\rangle=C\sb w\vert\theta\sb0\rangle=0.
\ee
If $\vert\theta\sb0\rangle$ is taken, we will not obtain any
additional coherent states because $C\sb
w\sp\dagger\vert\theta\sb0\rangle=0$, which imply that
$\vert\theta\sb0\rangle$ is invariant under the application of $D(z)$.
The only non-trivial possibility is to take
$\vert\varphi\sb0\rangle=\vert\theta\sb1\rangle$. However, the way in
which $[C\sb w,C\sb w\sp\dagger]$ acts on the basis vectors
$\vert\theta\sb n\rangle$ (see (3.15)) disables the factorization of
$D(z)$ to simplify the calculation of $D(z)\vert\theta\sb1\rangle$.
Therefore, we decided to consider the non-unitary operator
\be
D\sb w(z)\equiv e\sp{zC\sb w\sp\dagger},
\ee
and look for the states of the form:
\be
\vert z,w\rangle\sb d\propto D\sb w(z)\vert\theta\sb1\rangle.
\ee
Using (3.14), we obtain:
\be
\vert  z,w \rangle\sb d=\frac1{\sqrt{ \Gamma(w)\
{}\sb1F\sb1(w,1;r\sp2) }}\ \sum\sb{n=0}\sp{\infty} \frac{z\sp n}{n!}\
\sqrt{\Gamma(w+n)}\ \vert\theta\sb{n+1} \rangle .
\ee
Notice that, for $w=0$, we have $\vert  z,0 \rangle\sb d=
\vert\theta\sb{1} \rangle$, and there is no family of coherent states.

The completeness of this new set, $\lbrace \vert\theta\sb0\rangle$,
$\vert z,w\rangle\sb d$;  $z\in {\bf C}\rbrace$ , reads now:
\begin{eqnarray}
I &=& \vert\theta\sb{0}\rangle\langle\theta\sb{0}\vert + \int \vert
z,w \rangle\sb d\, {}\sb d\langle z,w\vert  \, d\mu\sb d(z,w)
\nonumber \\ & = & \vert\theta\sb{0}\rangle\langle\theta\sb{0}\vert +
\int \vert z,w\rangle\sb d\, {}\sb d\langle z,w\vert\, \sigma\sb
d(r,w)\, r \, dr\, d\varphi.
\end{eqnarray}
As the relevant values of $w$ are $w\geq 0$, we can define $\sigma\sb
d(r,w)=\Gamma(w)\, {}\sb1F\sb1(w,1;r\sp2)\eta(r\sp2,w)$, and following
a procedure similar to that of \cite{bg,bd}, it turns out that the
function $\eta(x,w)$ must satisfy:
\be
\int\sb0\sp\infty \eta(x,w)\ x\sp{m-1}\
dx=\frac{(\Gamma(m))\sp2}{\pi\, \Gamma(m+w-1)},\ \ m=1,2, \dots
\ee
We have to solve a typical ``momentum problem'' (see \cite{basu} an
references quoted therein). To do it, we can use the Mellin transform
technique, as we did to find $\sigma(r,w)$ in (4.8), and we get the
following result:
\begin{eqnarray}
\eta(x,w) &=&\frac1\pi \sum\sb{l=0}\sp{\infty} \frac{x\sp l}{(l!)\sp2\
\Gamma(w-l-1)} \ [-\ln x +2\psi(l+1)-\psi(w-l-1)],\ w\notin{\bf N};
\nonumber \\ \eta(x,n)& = & \frac1\pi \sum\sb{l=0}\sp{n-2} \frac{x\sp
l}{(l!)\sp2\ (n-l-2)!} \ [-\ln x +2\psi(l+1)-\psi(n-l-1)]   \nonumber
\\ & & +\, \frac{x\sp{n-1}}{\pi[(n-1)!]\sp2}\ {}\sb2F\sb2(1,1;n,n;-x)
,\quad n=1,2,3\dots,
\end{eqnarray}
where $\psi(y)=[\Gamma(y)]\sp{-1}d\Gamma(y)/dy$, and
${}\sb2F\sb2(1,1;n,n;x)$ is a generalized hypergeometric function
\cite{ba}. In the last ecuation, if $n=1$ the sum does not appear, and
the generalized hypergeometric function is very simple; the result is
$$
\eta(x,1)=e\sp{-x}/\pi.
$$

In the case we are considering, the reproducing kernel is
\be
{}\sb d\langle z,w \vert z',w \rangle\sb d = \frac{\sb1F\sb1(w,1;
\bar z z')}{\sqrt{\sb1F\sb1(w,1; r\sp2)  \ {}\sb1F\sb1(w,1;r'\sp2)}} .
\ee
The time evolution of these states is equal to that of (4.13),
$U(t)\vert z,w\rangle\sb d = e\sp{-i3t/2} \vert z(t),w\rangle\sb d$.
The mean value is defined as usual. Hence,
\be
\langle H\sb\lambda \rangle\sb d=\frac{3}{2}+ r\sp2\,S(w,r\sp2) ,
\quad \langle C\sb w\rangle\sb d= z\,S(w,r\sp2) , \quad \langle C\sb
w\sp\dagger\rangle\sb d= \bar{z} \, S(w,r\sp2) ,
\ee
where
\be
S(w,r\sp2)=w\, \frac{ {}\sb1F\sb1(w+1,2;r\sp2)}{
{}\sb1F\sb1(w,1;r\sp2)}.
\ee
{}From (3.16) one gets
\be
\langle X\sb w\rangle\sb d= \frac{(\bar z+z)}{\sqrt{2}} \, S(w,r\sp2)
, \qquad \langle P\sb w\rangle\sb d= \frac{i(\bar z-z)}{\sqrt{2}}\,
S(w,r\sp2) .
\ee
In order to obtain $\langle X\sb w\sp2\rangle\sb d$ and $\langle P\sb
w\sp2\rangle\sb d$, we find first
\be
\langle C\sb w\sp2\rangle\sb d =z\sp2\, T(w,r\sp2) , \qquad \langle
C\sb w\sp{\dagger 2}\rangle\sb d=\bar z\sp2\, T(w,r\sp2) ,
\ee
\be
\langle C\sb w C\sb w\sp\dagger + C\sb w\sp\dagger C\sb w \rangle\sb
d= -1+\frac{1 -w+ 2 w \,
{}\sb1F\sb1(w+1,1;r\sp2)}{{}\sb1F\sb1(w,1;r\sp2)} ,
\ee
where we have introduced the function $T(w,r\sp2)$, defined as
\be
T(w,r\sp2)=\frac{w(w+1)}{2}\, \frac{{}\sb1F\sb1(w+2,3;r\sp2)}
{{}\sb1F\sb1(w,1; r\sp2)}.
\ee
Therefore,
\be
\langle X\sb w\sp2\rangle\sb d=-\frac12+\frac12(\bar
z\sp2+z\sp2)\,T(w,r\sp2) + \frac{ 1 -w+ 2w\, {}\sb1F\sb1(w+1,1;r\sp2)
}{2\,{}\sb1F\sb1(w,1;r\sp2)} ,
\ee
\be
\langle P\sb w\sp2\rangle\sb d= -\frac12- \frac12(\bar
z\sp2+z\sp2)\,T(w,r\sp2) + \frac{ 1 - w+ 2w \,
{}\sb1F\sb1(w+1,1;r\sp2) }{2\, {}\sb1F\sb1(w,1;r\sp2)} .
\ee
Now, it is very easy to get the uncertainties $(\Delta X\sb w)\sb d, \
(\Delta P\sb w)\sb d$:
\begin{eqnarray}
(\Delta X\sb w)\sb d\sp2&=&\frac{(\bar z+z)\sp2}{2}\left[T(w,r\sp2) -
S\sp2(w,r\sp2)\right] -r\sp2\, T(w,r\sp2)  \nonumber \\ & & -\frac12
+\frac{1-w+ 2w \, {}\sb1F\sb1(w+1,1; r\sp2)} {2\,
{}\sb1F\sb1(w,1;r\sp2)} ,
\end{eqnarray}
\begin{eqnarray}
(\Delta P\sb w)\sb d\sp2&=&-\frac{(\bar z-z)\sp2}{2}\left[T(w,r\sp2) -
S\sp2(w,r\sp2)\right] -r\sp2\, T(w,r\sp2)  \nonumber \\ & &  -\frac12
+\frac{1-w+ 2w \, {}\sb1F\sb1(w+1,1;r\sp2)} {2\,
{}\sb1F\sb1(w,1;r\sp2)}.
\end{eqnarray}
{}From (4.42)-(4.43) it is clear that, in contrast to the previous case,
$(\Delta X\sb w)\sb d$,  $(\Delta P\sb w)\sb d$ and their product do
not have radial symmetry on the complex plane. Their dependence on
$\varphi={\rm arg}(z)$ means that, even though the evolution of a
coherent state of kind $\vert z,w\rangle\sb d$ is equal to the one of
a coherent state of kind $\vert z,w\rangle$, the uncertainties of
$X\sb w$ and $P\sb w$ change in time for the states $\vert
z,w\rangle\sb d$ but remain static for $\vert z,w\rangle$ (see
(4.21)). This immediately suggests a very interesting use of the
coherent states derived in this section: let us fix the values of $w$
and $r=\vert z\vert$ and let us take as initial condition one of the
states $\vert z,w\rangle\sb d$ having a maximum value of $(\Delta X\sb
w)\sb d$. From Figure 2 is is clear that this occurs for $\varphi=0$
or $\varphi=\pi$ if $0<w<1$ and for $\varphi=\pi/2$ or
$\varphi=3\pi/2$ if $w>1$; this can be also analytically proved. Now,
let us evolve this initial state a time $t=T/4$, where $T$ is the
period of the potentials (2.6), which in the units we are using
becomes $T=2\pi$. At the end of this interval the initial state has
evolved into a different coherent state, $\vert ze\sp{-
i\pi/2},w\rangle\sb d$, and the uncertainty $(\Delta X\sb w)\sb d$
will be minimum. This is nothing but a maximum efficiency squeezing
operation on the initial coherent state. The point is that we did not
have to apply any sophisticated sequence of external potentials on our
system to induce the squeezing operation. If we just select an
adequate coherent state $\vert z,w\rangle\sb d$ as the initial
condition, the evolution makes the work. Of course, it is possible the
design of a scheeme aimed to produce the inverse process, i.e., the
maximum efficiency expansion operation. It is up to the designer the
selection of which one of those processes he is interested in.

Finally, the harmonic oscillator limit of those states is
\be
\vert  z,w
\rangle\sb{d0}\equiv\lim\sb{\vert\lambda\vert\rightarrow\infty} \vert
z,w \rangle\sb{d} =\frac1{\sqrt{ \Gamma(w) \ {}\sb1F\sb1(w,1;r\sp2)}}
\sum\sb{n=0}\sp{\infty} \frac{z\sp n}{n!}\, \sqrt{\Gamma(w+n)}\ \vert
\psi\sb{n+1} \rangle.
\ee
Once again, we notice that $\vert  z,w \rangle\sb{d0}$ does not
coincide, in general, with a standard coherent state.

\section{Particular Cases}\label{sec5}

By taking three specific values of $w$, we analize now particular
situations for which our previous formulae take a simpler form. We
will study the cases with $w=0, \ w=1$ and $w=2$.

\subsection{The case $w=1$}

Here, the subspace ${\cal H}\sb r$, which is invariant under $C\sb1$
and $C\sb1\sp\dagger$, acquires also the property (3.3) of ${\cal
H}\sb s$. If we restrict the action of $C\sb1$ and $C\sb1\sp\dagger$
to ${\cal H}\sb r$, we get then a slight modification of the standard
representation of the Heisenberg algebra:
\be
C\sb1\vert\theta\sb n\rangle =  \sqrt{n-1}\ \vert\theta\sb{n-
1}\rangle, \quad C\sb1\sp\dagger \vert\theta\sb n\rangle = \sqrt{n}\
\vert\theta\sb{n+1} \rangle ,\quad
[C\sb1,C\sb1\sp\dagger]\vert\theta\sb n\rangle = \vert\theta\sb
n\rangle, \quad n\geq 1.
\ee
The two sets of coherent states derived in Section 4 become equal:
\be
\vert z,1\rangle=\vert z,1\rangle\sb d=e\sp{-
r\sp2/2}\sum\sb{n=0}\sp\infty \frac{z\sp
n}{\sqrt{n!}}\vert\theta\sb{n+1}\rangle.
\ee
They are the standard coherent states if we relabel the eigenstates of
$H\sb\lambda$ as $\vert\varphi\sb n\rangle\equiv
\vert\theta\sb{n+1}\rangle$. The measure functions $\sigma(r,1), \
\sigma\sb d(r,1)$, and the kernels (4.10), (4.33) are transformed into
the standard ones:
\be
\sigma(r,1)=\sigma\sb d(r,1)=\frac1{\pi}, \qquad \langle z,1\vert
z',1\rangle = {}\sb d\langle z,1\vert z',1\rangle\sb d=\exp\left(-
\frac{r\sp2}2 -\frac{r'\sp2}2+\bar zz'\right).
\ee
Due to the fact that the state on ${\cal H}\sb r$ with the minimum
value of the energy is $\vert\theta\sb1 \rangle$, $\langle
H\sb\lambda\rangle$ becomes slightly different of the standard result:
\be
\langle H\sb\lambda\rangle =\langle H\sb\lambda\rangle\sb
d=\frac32+r\sp2.
\ee
However, those coherent states are minimum uncertainty states, as they
verify
\be
(\Delta X\sb1)(\Delta P\sb1) = (\Delta X\sb1)\sb d(\Delta P\sb1)\sb
d=\frac12.
\ee
This result justifies, once again, the name selected for the algebra
generated by $C\sb w$ and $C\sb w\sp\dagger$, because we have found
one $w$-value for which it reduces to the standard Heisenberg algebra
on ${\cal H}\sb r\subset {\cal H}$. We will see next that $w=1$ is not
the only value inducing this behaviour.

\subsection{The case $w=0$}

Let us take the limit $w\rightarrow 0$ in all formulae of Sections 3
and 4. The subspace ${\cal H}\sb r$ decomposes now into two invariant
subspaces: one of them is generated by $\vert\theta\sb1\rangle$ while
the other one is ${\cal H}\sb s$, generated by $\lbrace
\vert\theta\sb n\rangle, \ n\geq2\rbrace $. The relevant
representation of $C\sb0$ and $C\sb0\sp\dagger$ arises from the
restriction of these operators to ${\cal H}\sb s$. We get again a
slight modification of the standard Heisenberg algebra representation:
\be
C\sb0\vert\theta\sb n\rangle = \sqrt{n-2}\ \vert\theta\sb{n-1}\rangle,
\quad C\sb0\sp\dagger \vert\theta\sb n\rangle = \sqrt{n-1}\
\vert\theta\sb{n+1}\rangle , \quad
[C\sb0,C\sb0\sp\dagger]\vert\theta\sb n\rangle = \vert\theta\sb
n\rangle, \quad n\geq2.
\ee
The coherent states which are eigenstates of $C\sb0$, denoted as
$\vert z,0\rangle$, are given in (4.6). However, those arising of the
action of $D\sb0(z)=\exp(zC\sb0\sp\dagger)$, denoted as $\vert
z,0\rangle\sb d$, cannot be found from (4.29) because those were
derived taking $\vert \theta\sb1\rangle$ as extremal state, but now it
does not belong to ${\cal H}\sb s$. Here, the extremal state inducing
non trivial coherent states is $\vert\theta\sb2\rangle$. A similar
calculation as that of subsection 4.2 leads immediately to $\vert
z,0\rangle\sb d$. These states, modulo a phase, are equal to those
obtained in (4.6):
\be
\vert z,0\rangle\sb d = e\sp{-i\varphi} \vert z,0\rangle = e\sp{-
r\sp2/2}\sum\sb{n=0}\sp\infty \frac{z\sp n}{\sqrt{n!}}\
\vert\theta\sb{n+2}\rangle.
\ee
They are again as the standard coherent states. The measure function
(4.8) and the kernel (4.10) are equal (modulo a phase) to the standard
ones, and to those of the previous section (see (5.3)). The mean value
of $H\sb\lambda$ is different because we depart from a different
extremal state:
\be
\langle H\sb\lambda\rangle =\langle H\sb\lambda\rangle\sb d=\frac52 +
r\sp2.
\ee
However, once again we find that $\vert z,0\rangle$ are minimum
uncertainty states:
\be
(\Delta X\sb0)(\Delta P\sb0) = (\Delta X\sb0)\sb d(\Delta P\sb0)\sb
d=\frac12.
\ee
Thus, we have found some additional information which we did not
foresee before: through the analysis of the coherent states resulting
from the two definitions considered in Section 4, we have been able to
construct the coherent states characteristic of the third definition.
We will analize next the simplest particular case involving a
representation qualitatively different from the standard Heisenberg
algebra representation.

\subsection{The case $w=2$}

Let us put $w=2$ in all the relationships of Sections 3 and 4. As in
the general case of Section 3, the relevant subspace is ${\cal H}\sb
r$, generated by the basis $\lbrace \vert\theta\sb n\rangle, n\geq
1\rbrace $. However, unlike the two previous particular cases, we do
not obtain now a representation of the standard Heisenberg algebra,
but
\be
C\sb2\vert\theta\sb n\rangle = (1-\delta\sb{n,1})\ \sqrt{n}\
\vert\theta\sb{n-1}\rangle,      \quad C\sb2\sp\dagger \vert\theta\sb
n\rangle = \sqrt{n+1}\ \vert\theta\sb{n+1}\rangle ,
\ee
\be
[C\sb2,C\sb2\sp\dagger]\vert\theta\sb n\rangle = (1+\delta\sb{n,1})
\vert\theta\sb n\rangle =\left\lbrace  \begin{array}{ll}
2\vert\theta\sb n\rangle, & n=1; \\ \vert\theta\sb n\rangle, & n\geq 2
. \end{array} \right.
\ee
This difference is the reason that the two sets of coherent states
considered in Section 4 are not equal:
\be
\vert z,2\rangle = \frac{r}{\sqrt{e\sp{r\sp2}-1}}\
\sum\sb{n=0}\sp{\infty} \frac{z\sp n}{\sqrt{(n+1)!}} \
\vert\theta\sb{n+1}\rangle,
\ee
\be
\vert z,2\rangle\sb d = \frac{e\sp{-{r\sp2}/2}}{\sqrt{1+r\sp2}}
\sum\sb{n=0}\sp{\infty} z \sp n \sqrt{\frac{(n+1)}{n!}}\
\vert\theta\sb{n+1}\rangle.
\ee
The measure functions and kernels are also different for the two
families:
\be
\sigma(r,2)= \frac{1-e\sp{-r\sp2}}{\pi}, \qquad \sigma\sb d(r,2)=
\frac{1}{\pi}\, e\sp{r\sp2}\,(1+r\sp2)\,{\rm E}\sb1(r\sp2),
\ee
\be
\langle z,2 \vert z',2 \rangle =e\sp{i(\varphi-\varphi')}\
\frac{e\sp{\bar z z' }-1} {\sqrt{(e\sp{r\sp2}-1) (e\sp{r'\sp2}-1)}},
\ee
\be
\quad {}\sb d\langle z,2 \vert z',2 \rangle\sb d =\frac{(1+\bar
zz')}{\sqrt{(1+r\sp2)(1+r'\sp2)}}\ \exp\left(-\frac{r\sp2}2 -
\frac{r'\sp2}2 +\bar zz'\right),
\ee
where ${\rm E}\sb1(x)$ is the exponential integral function. As we
could expect, the mean values of $H\sb\lambda$ in both sets are not
equal:
\begin{eqnarray}
\langle z,2\vert H\sb\lambda \vert z,2 \rangle  &=& \frac12
+\frac{r\sp2 }{1 - e\sp{-r\sp2}},
\end{eqnarray}
\begin{eqnarray}
{}\sb d\langle z,2\vert H\sb\lambda \vert z,2 \rangle\sb d  &=&
\frac32 +\frac{2+ r\sp2}{1 + r\sp2}\ r\sp2.
\end{eqnarray}
Finally, it turns out that the uncertainties of $X\sb2$, $P\sb2$ and
their product are distinct on both sets:
\be
(\Delta X\sb2)\sp2 = (\Delta P\sb2)\sp2= (\Delta X\sb2)(\Delta
P\sb2)=\frac12\left(1+ \frac{r\sp2}{e\sp{r\sp2}-1}\right),
\ee
\be
(\Delta X\sb2)\sb d\sp2=\frac1{2(1+r\sp2)}\left[ 3 + r\sp2 - e\sp{-
r\sp2} - \frac{(\bar z+z)\sp2}{1+r\sp2}\right],
\ee
\be
(\Delta P\sb2)\sb d\sp2=\frac1{2(1+r\sp2)}\left[ 3 + r\sp2 - e\sp{-
r\sp2} + \frac{(\bar z-z)\sp2}{1+r\sp2}\right],
\ee
\be
(\Delta X\sb2)\sb d(\Delta P\sb2)\sb d=\frac1{2(1+r\sp2)}\left[\left(3
+ r\sp2 - e\sp{-r\sp2}\right)\left(\frac{3+r\sp4}{1+r\sp2} - e\sp{-
r\sp2}\right) - \left(\frac{\bar z\sp2-z\sp2}{1+r\sp2}
\right)\sp2\right]\sp{1/2}.
\ee
Let us notice, once again, that the uncertainties of $X\sb2$ and
$P\sb2$ on the coherent states $\vert z,2\rangle\sb d$ do not have
radial symmetry (see (5.20)-(5.22)). Then, for these states it is
possible the design of a prescription to induce a maximum efficiency
squeezing operation by means of the natural evolution of the system,
as it was discussed at the end of Section 4.

\medskip

Until now, all our results are concerned with the intrinsic structure
of $H\sb\lambda$ and the distorted coordinate and momentum operators
appropiate to this structure. Nevertheless, it would be interesting to
find the dispersion for the standard coordinate $X$ and momentum $P$
in the coherent states here derived. This  is hard to do for $w$ and
$\lambda$ arbitrary, but can be easily performed in the harmonic
oscillator limit, and for particular values of $w$. We will restrict
ourselves to the case $\vert\lambda\vert\rightarrow\infty$ and $w=2$.
This is justified because, in this limit, $C\sb w$ and $C\sb
w\sp\dagger$ behave on ${\cal H}\sb r$ almost exactly as the usual
annihilation and creation operators do on ${\cal H}$:
\be
C\sb{2,0}\vert\psi\sb n\rangle\equiv \lim\sb{\vert
\lambda\vert\rightarrow\infty} C\sb2\vert\theta\sb n\rangle =
\left\lbrace  \begin{array}{ll}  0, & n=1 ; \\ a \vert\psi\sb n\rangle
= \sqrt{n}\, \vert\psi\sb{n-1} \rangle, & n \geq 2; \end{array}
\right.
\ee
\be
C\sb{2,0}\sp\dagger\vert\psi\sb n\rangle\equiv \lim\sb{\vert
\lambda\vert\rightarrow\infty} C\sb2\sp\dagger \vert\theta\sb n\rangle
= a\sp\dagger\vert\psi\sb n\rangle = \sqrt{n+1}\, \vert\psi\sb{n+1}
\rangle, \quad n \geq 1.
\ee
All the results derived for $X\sb2, \ P\sb2, \ \vert z,2\rangle$ and
$\vert z,2\rangle\sb{d}$ remain valid in this limit, where we denote
$\vert z,2\rangle\sb0, \ \vert z,2\rangle\sb{d0}$  the coherent states
after the limit, and $(\Delta X\sb2)\sb0, \ (\Delta X\sb2)\sb{d0}$
the uncertainties of $X\sb2$ on both sets with the same notation for
any other operator. The uncertainties we will obtain for $X$ and $P$
will be compared with those for $X\sb{2}$ and $P\sb{2}$ in
(5.19)-(5.22).

First, the mean values of $X=(a\sp\dagger + a)/\sqrt{2}$ and
$P=i(a\sp\dagger- a)/\sqrt{2}$ in the states $\vert z,2\rangle\sb0$
are:
\be
{}\sb{0}\langle z,2\vert X \vert z,2 \rangle\sb{0}  = \frac{\bar z+ z
}{\sqrt{2}},  \qquad \quad {}\sb{0}\langle z,2\vert P \vert z,2
\rangle\sb{0}=i\ \frac{\bar z  - z }{\sqrt{2}}.
\ee
We evaluate also
\be
{}\sb{0}\langle z,2\vert a a\sp\dagger + a\sp\dagger a \vert z,2
\rangle\sb{0}  = 1 + \frac{2r\sp2}{1-e\sp{-r\sp2}};
\ee
\be
{}\sb{0}\langle z,2\vert a\sp2 \vert z,2 \rangle\sb{0}  =
\overline{{}\sb{0}\langle z,2\vert a\sp{\dagger 2} \vert z,2
\rangle\sb{0}}=z\sp2.
\ee
Hence,
\begin{eqnarray}
{}\sb{0}\langle z,2\vert X\sp2 \vert z,2 \rangle\sb{0}
&=&\frac12\left(1+z\sp2+{\bar z }\sp2 +\frac{2r\sp2}{1-e\sp{-r\sp2}}
\right); \\   {}\sb{0}\langle z,2\vert P\sp2 \vert z,2 \rangle\sb{0}
&=&\frac12\left(1-z\sp2-{\bar z }\sp2 +\frac{2r\sp2}{1-e\sp{-r\sp2}}
\right).
\end{eqnarray}
We get now the dispersion of $X$ and $P$ and their product:
\be
(\Delta X)\sb0\sp2= (\Delta P)\sb0\sp2=(\Delta X)\sb0(\Delta P)\sb0=
\frac12+ \frac{r\sp2}{e\sp{r\sp2}-1}.
\ee
For $\vert z,2\rangle\sb{d0}$ we have:
\be
{}\sb{d0}\langle z,2 \vert X  \vert  z,2 \rangle\sb{d0}=
\left(\frac{2+r\sp2}{1+r\sp2}\right)\frac{(\bar z+z)}{\sqrt 2} ,
\ee
\be
{}\sb{d0}\langle z,2 \vert P  \vert  z,2 \rangle\sb{d0}=
\left(\frac{2+r\sp2}{1+r\sp2}\right)\frac{i(\bar z-z)}{\sqrt 2} .
\ee
In order to evaluate easily the deviations of $X$ and $P$, we find
first
\be
{}\sb{d0}\langle z,2 \vert aa\sp\dagger + a\sp\dagger a  \vert
z,2\rangle\sb{d0}=\frac{(r\sp2+3)(2r\sp2+1)}{1+r\sp2} ,
\ee
\be
{}\sb{d0}\langle z,2 \vert a\sp2  \vert  z,2\rangle\sb{d0}=
\overline{{}\sb{d0}\langle z,2 \vert a\sp{\dagger 2}  \vert
z,2\rangle\sb{d0}}= \left(\frac{3+r\sp2}{1+r\sp2}\right) z\sp2.
\ee
Hence,
\be
{}\sb{d0}\langle z,2 \vert X\sp2  \vert  z,2\rangle\sb{d0} =
\frac12\left(\frac{r\sp2+3}{r\sp2+1}\right)\left[(\bar z+z)\sp2+1
\right],
\ee
\be
{}\sb{d0}\langle z,2 \vert P\sp2  \vert  z,2\rangle\sb{d0} =
\frac12\left(\frac{r\sp2+3}{r\sp2+1}\right)\left[-(\bar z-z)\sp2+1
\right].
\ee
Using the previous results, the dispersions of $X$ and $P$ are given
by:
\be
(\Delta X)\sb{d0}\sp2= \frac12 \left(\frac{r\sp2+3}{r\sp2+1}\right) -
\frac12\left(\frac{\bar z+z}{r\sp2+1} \right)\sp2,
\ee
and
\be
(\Delta P)\sb{d0}\sp2= \frac12 \left(\frac{r\sp2+3}{r\sp2+1}\right) +
\frac12\left(\frac{\bar z-z}{r\sp2+1} \right)\sp2.
\ee
A plot of $[(\Delta X)\sb{d0}(\Delta P)\sb{d0}]$ as a function of $z$
is shown in Figure 3. By comparing equations (5.30) and (5.37)-(5.38)
with (5.19)-(5.21), we realize that both results are different. The
reason is that the subspace ${\cal H}\sb r$, which is invariant under
$X\sb2, \ P\sb2$, is not invariant under $X, \ P$. This fact induces
additional terms in the formulas, which produce the final difference
between the deviations for $X, \ P$ and for $X\sb2, \ P\sb2$.

\section{Concluding Remarks}\label{sec6}

We have been able to answer the two questions posed in the
Introduction: there exist creation and annihilation operators for
$H\sb\lambda$, the ones with $w=0$ and $w=1$, which behave as the
generators of the Heisenberg algebra if their action is rectricted to
the invariant subspaces ${\cal H}\sb s$ and ${\cal H}\sb r$
respectively; the two sets of coherent states associated to each one
of these operators became essentially equal to the standard coherent
states of the harmonic oscillator. It is important to remind that
these coherent states turned out to be minimum uncertainty states for
the distorted coordinate and momentum operators; therefore, we were
able to construct indirectly the coherent states described in the
third definition. If we had restricted our considerations just to
answer those questions, we would never have found the rich family of
annihilation and creation operators of the distorted Heisenberg
algebra characteristic of $H\sb\lambda$. Moreover, we would never have
found the coherent states $\vert z,w\rangle\sb d$ on which it is
possible to induce the maximum efficiency squeezing operation.

We would like to end this paper with a short comment concerning the
widely discussed geometric phase \cite{sw,fn}. It arises for any state
$\vert \psi(t)\rangle$ which evolves in a cyclic way during a time
interval $[0,\tau]$, i.e., $\vert\psi(\tau)\rangle=
e\sp{i\phi}\vert\psi(0)\rangle, \quad \phi\in{\bf R}$. The phase
$\phi$ can be decomposed as a dynamic plus a geometric part, the last
one named geometric phase and denoted $\beta$. If the evolution is
induced by the Hamiltonian $H(t)$, it turns out that $\beta$ takes the
form \cite{sw} ($\hbar=1$):
\be
\beta=\phi+\int\sb0\sp\tau \langle\psi(t)\vert H(t)\vert\psi(t)
\rangle dt.
\ee
For a time independent Hamiltonian with equally spaced discrete
spectrum, any initial state evolves cyclically \cite{fe}. This is true
for our family of Hamiltonians (2.5)-(2.6) and the coherent states of
Section 4. They are cyclic with period $\tau=2\pi$, $U(2\pi)\vert
z,w\rangle=e\sp{-i3\pi} \vert z,w\rangle$ and $U(2\pi)\vert
z,w\rangle\sb d=e\sp{-i3\pi} \vert z,w\rangle\sb d$. A direct
calculation leads us to the following expressions for the geometric
phases:
\begin{eqnarray}
\beta&=&2\pi\left(\langle H\sb\lambda\rangle -\frac32\right)=
2\pi\left(\frac{ {}\sb1F\sb1(2,w; r\sp2) }{ {}\sb1F\sb1(1,w; r\sp2)} -
1\right), \\ \beta\sb d&=&2\pi\left(\langle H\sb\lambda\rangle\sb d -
\frac32\right)= 2\pi \, r\sp2 S(w,r\sp2).
\end{eqnarray}
Hence, $\langle H\sb\lambda\rangle$ and $\langle H\sb\lambda\rangle\sb
d$ are, essentially, geometric quantities in the same sense as $\beta$
and $\beta\sb d$ are geometric \cite{fe}.

\section{Acknowledgements}\label{}

The authors acknowledge CONACYT (M\'exico) for finantial support.
L.M.N. also thanks the ``Centre de Recherches Math\'ematiques''
(Montr\'eal) and CINVESTAV (M\'exico) for  kind hospitality, as well
as DGICYT (Spain) for a fellowship and partial support (project
PB92-0255).

\vfill\eject


\vfill\eject

\
\bigskip\bigskip

\centerline{\large\bf Figure Captions}

\bigskip\bigskip

\noindent {\bf Figure 1.} Plot of $(\Delta X\sb w)(\Delta P\sb w)$ as
a function of $r=\vert z\vert$ for different values of $w$.

\bigskip

\noindent {\bf Figure 2.} Plot of $(\Delta X\sb w)\sb d$ as a function
of $z$ for two $w$-values: a) $w=0.5$; b) $w=5$. Notice that, for
fixed values of $r=\vert z\vert$ and $w$, the maximum of $(\Delta X\sb
w)\sb d$ occurs for $\varphi={\rm arg}(z)=0$ or $\pi$ if $0<w<1$,
while it occurs for $\varphi=\pi/2$ or $3\pi/2$ if $w>1$. The
corresponding graphs for $(\Delta P\sb w)\sb d$ can be obtained from
the previous ones by a rotation of $\pi/2$ around the vertical axis.

\bigskip

\noindent {\bf Figure 3.} Plot of the uncertainty product for the
position $X$ and momentum $P$ operators as a function of $z$ in the
harmonic oscillator limit for the coherent states $\vert z,2\rangle\sb
d$.

\end{document}